\documentclass[fleqn,usenatbib]{mnras}

\usepackage[T1]{fontenc}

\DeclareRobustCommand{\VAN}[3]{#2}
\let\VANthebibliography\thebibliography
\def\thebibliography{\DeclareRobustCommand{\VAN}[3]{##3}\VANthebibliography}

\hypersetup{linkcolor=blue,citecolor=blue,filecolor=cyan,urlcolor=magenta}
\usepackage{graphicx}   
\usepackage{amsmath}    
\usepackage{amssymb}    
\usepackage{newtxtext,newtxmath}
\usepackage{color}
\usepackage{threeparttable}
\usepackage{hyperref}

\def\mr{\mathrm}

\newcommand{\lta}{\lower 2pt \hbox{$\, \buildrel {\scriptstyle <}\over {\scriptstyle \sim}\,$}}
\newcommand{\gta}{\lower 2pt \hbox{$\, \buildrel {\scriptstyle >}\over {\scriptstyle \sim}\,$}}
\definecolor{blazeorange}{rgb}{1.0, 0.4, 0.0}
\definecolor{seagreen}{rgb}{0.18, 0.55, 0.34}
\definecolor{rufous}{rgb}{0.66, 0.11, 0.03}
\definecolor{royalfuchsia}{rgb}{0.79, 0.17, 0.57}
\definecolor{scarlet}{rgb}{1.0, 0.13, 0.0}
\definecolor{royalpurple}{rgb}{0.47, 0.32, 0.66}

\title[Hybrid model for FRBs]{Hybrid Pulsar-Magnetar Model for FRB 20191221A}

\author[Beniamini \& Kumar]{
	Paz Beniamini$^{1,2}$\thanks{pazb@openu.ac.il}\& Pawan Kumar$^3$\thanks{pk@astro.as.utexas.edu}\\
	$^{1}$Department of Natural Sciences, The Open University of Israel, P.O Box 808, Ra'anana 4353701, Israel\\
	$^{2}$Astrophysics Research Center of the Open university (ARCO), The Open University of Israel, P.O Box 808, Ra'anana 4353701, Israel\\
	$^3$Department of Astronomy, University of Texas at Austin, Austin, TX 78712, USA
}

\begin{document}
	\label{firstpage}
	\pagerange{\pageref{firstpage}--\pageref{lastpage}}
	\maketitle
	
	\begin{abstract}
		We show that the 216.8$\pm$0.1 ms periodicity reported for the fast radio burst (FRB) 20191221A is very constraining for burst models. The high accuracy of burst periodicity (better than one part in 10$^3$), and the 2\% duty cycle (ratio of burst-duration and inter-burst interval), suggest a pulsar-like rotating beam model for the observed activity; the radio waves are produced along open field lines within $\sim 10^7$ cm of the neutron star surface, and the beam periodically sweeps across the observer as the star spins. According to this picture, FRB 20191221A is a factor $\sim 10^{12}$ scaled up version of galactic pulsars with one major difference: whereas pulsars convert rotational kinetic energy to EM waves, the outbursts of 20191221A require conversion of magnetic energy to radiation.
		
	\end{abstract}

	
	
	
	\section{Introduction}
	\label{sec:intro}
	The fast radio burst (FRB) 20191221A is a non-repeater with a multi-component lightcurve detected by CHIME \cite{subsecPCHIME}. The overall duration of the burst was $t_{\rm FRB}\!\sim\! 3$\, s. The spacing of the peaks is reported to be highly periodic with a period of $P\!=\!216.8 \pm 0.1 \mbox{ ms}$ (and with some null periods). The individual pulses are narrow, corresponding to a duty cycle of $\eta \!\sim\! 0.018$. A lower limit on the fluence of the burst is $\mathcal{F}\!>\!1.2\times 10^3\mbox{Jy ms}$ and a lower limit on the peak flux is $F_{\rm p}\!>\!2\,$Jy. Its reported dispersion measure is $\mbox{DM}\!=\!367\, \mbox{pc cm}^{-3}$. The maximum contribution to the DM from the Milky Way in the direction of the burst is estimated as $\sim \!90\,\mbox{pc cm}^{-3}$. Taking, conservatively, the host galaxy DM to be $\sim \!100\,\mbox{pc cm}^{-3}$ we get $\mbox{DM}_{\rm IGM}\gtrsim 180\,\mbox{pc cm}^{-3}$, leading to a lower limit on the distance $d_{\rm L}\!\gtrsim\! 400$\,Mpc. Thus, the isotropic equivalent luminosity in the 400-800 MHz band is $\sim \!2\!\times\! 10^{41}$\,erg s$^{-1}$ and the isotropic energy release in the radio band $\sim\! 10^{41}$\,erg. 
	
	In the last few years, there have been multiple lines of argument connecting FRBs with neutron star progenitors \citep{Masui+15,Katz16,Kumar+17,Metzger+17, Margalit&Metzger18,Metzger+19, KumarBosnjak2020,Michilli+18,Wadiasingh2019,Heintz2020,CHIME2020,Li2020,STARE2020,LBK2022}.
	Considering a NS central engine, two possible mechanisms for explaining the short duration periodicity seen in FRB 20191221A are: (i) the NS spin and (ii) crustal oscillations. Outside of the light-cylinder, the magnetic field lines are carried by the outflow from the NS and they are no longer in causal contact with the surface of the NS. Therefore, they generally cannot reflect the periodicity associated with the NS spin \footnote{ We have assumed here that the emitting particles' trajectories are bent according to the NS's field lines. This is true as long as the Larmour radius of the particles at the ejection radius, $R_{\rm L}(r)$ is smaller than the ejection radius, $r$. For a dipolar field structure this condition is satisfied for $r<r_{\rm esc}\equiv 8\times 10^{13}(B/10^{15})^{1/2}(\gamma/100)^{-1/2}\mbox{ cm}$ where $B$ is the dipole field strength at the NS surface and $\gamma$ is the emitting particles' Lorentz factor (LF). We see that $r_{\rm esc}\gg R_{\rm LC}=10^9\mbox{ cm}$ (where $R_{\rm LC}\equiv c/\Omega$) for values of $B,\gamma$ that are of relevance in the FRB context (the opposite situation has been explored in the context of pulsar $\gamma$-ray radiation, see \citealt{Cerutti2016,Philippov2018,Kalapotharakos2018}). Indeed, taking into account the observed duty cycle, the emitting radius must be $r_{\rm em}\lesssim 10R_0$ (see \S \ref{sec:constraints}) which makes the relevant limiting condition much stricter than implied above, i.e. $r_{\rm esc}<r_{\rm em}$. Finally, we note that even if this condition were to be satisfied, a scenario such as this would imply that even a small change in the ejection radius would cause significant changes in the accuracy of the observed periodicity, in contrast with observations.}. An exception to this rule is the case of a striped wind outflow \citep{Coroniti1990,LyubarskyKirk2001,DS2002,Giannios2006},
	in which a high magnetization outflow can carry the imprint of spin periodicity to radii outside the light-cylinder in the form of magnetic fields with reversing polarity. These oppositely oriented stripes reconnect a large distance from the source, and each reconnection event may lead to a spike in the observed lightcurve.
	If there is a fixed radius at which stripes reconnect then the spin period can be directly imprinted on the observed lightcurve
	in this fashion \footnote{A variant of the reconnection model, considers forced reconnection. As discussed in \S \ref{sec:forcedrec}, in that scenario the temporal separation between radio spikes corresponds to the frequency of the flare ejections (which, if periodic, could originate from the crustal oscillation frequency), {\it not} to the spin period.}
	. However, the reconnection velocity is estimated to be sub-relativistic, $\beta_{\rm rec}\!\sim\! 0.1\!-\!0.25$ \citep{Lyubarsky2005}, and as a result the duration of each spike in the lightcurve is $t_{\rm F}\!\sim \! P/\beta>P$, i.e {\it larger} than the separation between spikes \citep{BG2016}. If the plasma accelerated at reconnection sites outflows with Lorentz factor $\Gamma'$ relative to the co-moving frame of the striped wind, then $t_{\rm F}$ can be shortened by $\Gamma'$ \citep{BG2016}. However $\Gamma'$ needs to be unrealistically large to get the observed $t_{\rm F}/P\!\sim\! 1/50$ (requiring in turn a huge magnetization which is unfavourable for synchrotron maser emission). Similarly, if the spatial extent of the region from which matter flows towards the reconnection layer is much smaller than the distance between stripes in the wind, then $t_{\rm F}$ can also be reduced. However, this requires fine-tuned geometry and typically results in a significant loss of efficiency of converting outflow energy to observed radiation \citep{BK2020}. We conclude that if the origin of the observed (low duty cycle) periodicity were to be the NS spin period, then the radio emission must be produced inside the NS light-cylinder within its magnetospheric. If, instead, the observed periodicity is driven by crustal oscillations, then the coherent radio waves could be produced either in the magnetosphere or outside the light-cylinder, i.e. no constraint can be placed on the distance where the radiation is produced in this case.
	
	We begin by exploring the general implications of associating the observed periodicity with crustal oscillations (\S \ref{sec:crustal}). We then discuss specific further constraints on far-away models of coherent radio emission (\S \ref{sec:FarAway}). In \S \ref{sec:constraints} we turn to discuss the parameter space for nearby models in which $P$ reflects the underlying rotation of the NS. We discuss some consequences of this picture and conclude in \S \ref{sec:conc}.
	
	\section{Crustal oscillations}
	\label{sec:crustal}
	If the underlying object producing FRB 191221A is a magnetar then the measured period $P\!=\!217\mbox{ ms}$ may be associated with crustal oscillations as suggested by \cite{subsecPCHIME}. Some insight on this possibility can be gained by comparison to known Galactic magnetars.
	QPOs (thought to result from crustal oscillations) have been detected in SGR bursts and giant flares from Galactic magnetars with frequencies of $\sim \!18\!-\!1800$\,Hz \citep{Watts2016}. While a lower frequency of $5$\,Hz cannot be ruled out as due to crustal oscillations, it must be relatively rare compared to $\sim \! 100$\,Hz oscillations based on the extensive QPO data \citep{Israel2005,Strohmayer2005,Watts2006,Huppenkothen2013,Watts2016,Miller2019}.
	Moreover, if crustal oscillations were responsible for producing FRBs then why is it that the radio signal is concentrated to 4 ms while the crustal oscillation period is $\sim 50$ times longer? Indeed, observationally, QPOs typically have order unity duty cycles \citep{Huppenkothen2013}.
	Furthermore, since 5 Hz is a high-order overtone of the characteristic $\sim 100$\,Hz crustal frequency, it is inexplicable that more harmonics are not present in the data thereby making the extremely accurate periodicity found in the CHIME data ($\Delta P/P < 10^{-3}$) quasi-period or much less accurate. An additional difficulty for the crustal oscillation scenario for FRB periodicity, if generic, is that higher frequency QPOs should in fact be easier to detect in non-repeating FRBs considering that there are $\sim 15$ times more bursts whose durations are $t_{\rm FRB}\gg 1/(100\mbox{ Hz})\sim 10$\,ms than bursts with $t_{\rm FRB}\gg1/(5\mbox{ Hz})\sim 200$\,ms. Thus, longer crustal oscillations periods ($> 200$ms) are not only less frequent but also detectable in a much smaller pool of non-repeating bursts. These arguments make clear that the interpretation of FRB 191221A's period as due to crustal oscillations poses a number of serious challenges.
	
	\section{Far away models}
	\label{sec:FarAway}
	We discuss in this section whether the observed $216.8\pm 1$ ms periodicity for the bursts of FRB 20191221A is consistent with the class of the models where the radio waves are produced outside the NS light-cylinder by relativistic outflows that are periodically produced by a magnetar. This class of models where the radio waves are produced at a large distance from the NS are referred to as the {\it Far away Model.} We discuss below four possible scenarios for the far away model. One of these is where a fast moving outflow collides with a slower moving outflow that was launched a little earlier, and maser emission is generated in the shocked region of the colliding outflows. This is known as the {\it internal shock} model. The second model we discuss is where outflows interact with the wind from the magnetar that is launched in between the outflow emission episodes. The third possibility is that the radio waves are produced in the external shock which is refreshed by the energy deposited by successive outflows when they catch up with the decelerating external shock. The final possibility, is that of forced reconnection where following a flare event, a magnetic pulse travels outwards from the magnetar and forces stripes of oppositely oriented magnetic field lines in the pulsar wind to reconnect above the light-cylinder. %
	
	\subsection{Reproducing observed periodicity}
	\subsubsection{Internal shocks}
	Far-away emission may arise from collisions between relativistic shells ejected by the central engine. In such a scenario, each pulse in the lightcurve corresponds to a collision between two such shells. As we show below, even if the shells are ejected precisely periodically, the observed burst times will not preserve this underlying periodicity.
	
\begin{figure}
		\centering
		\includegraphics[width = 0.44\textwidth]{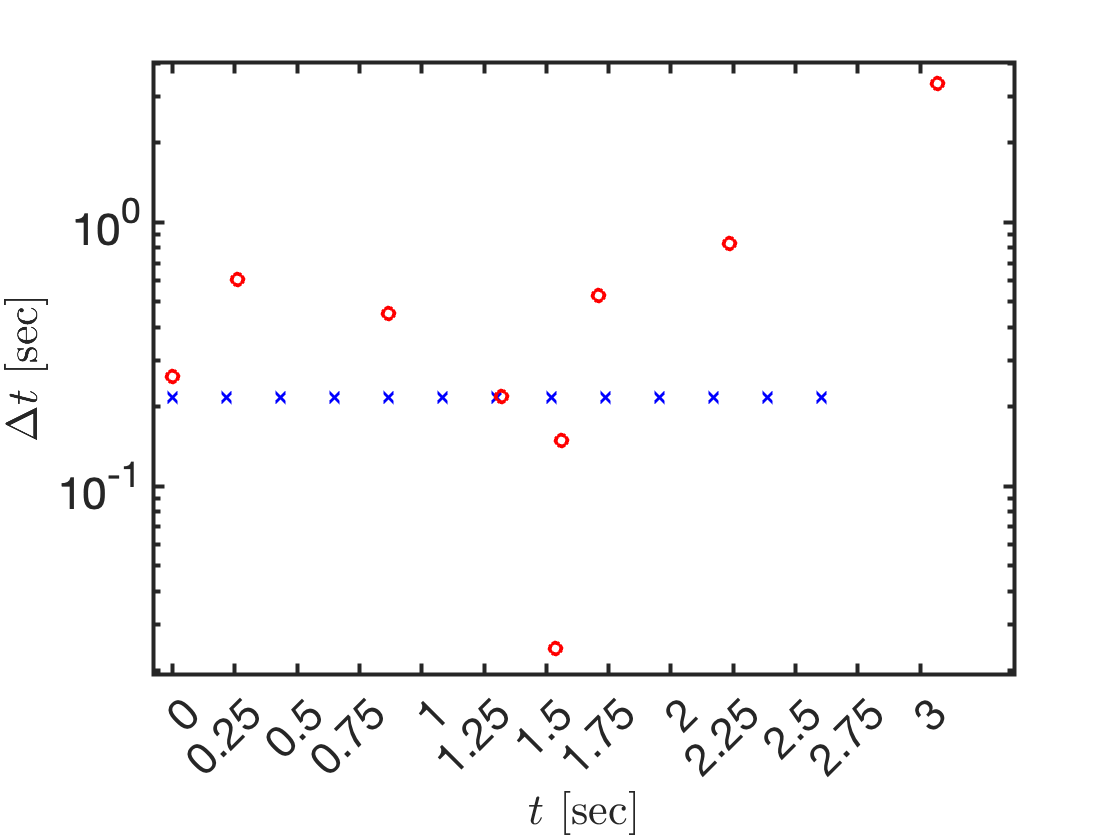}
		\includegraphics[width = 0.44\textwidth]{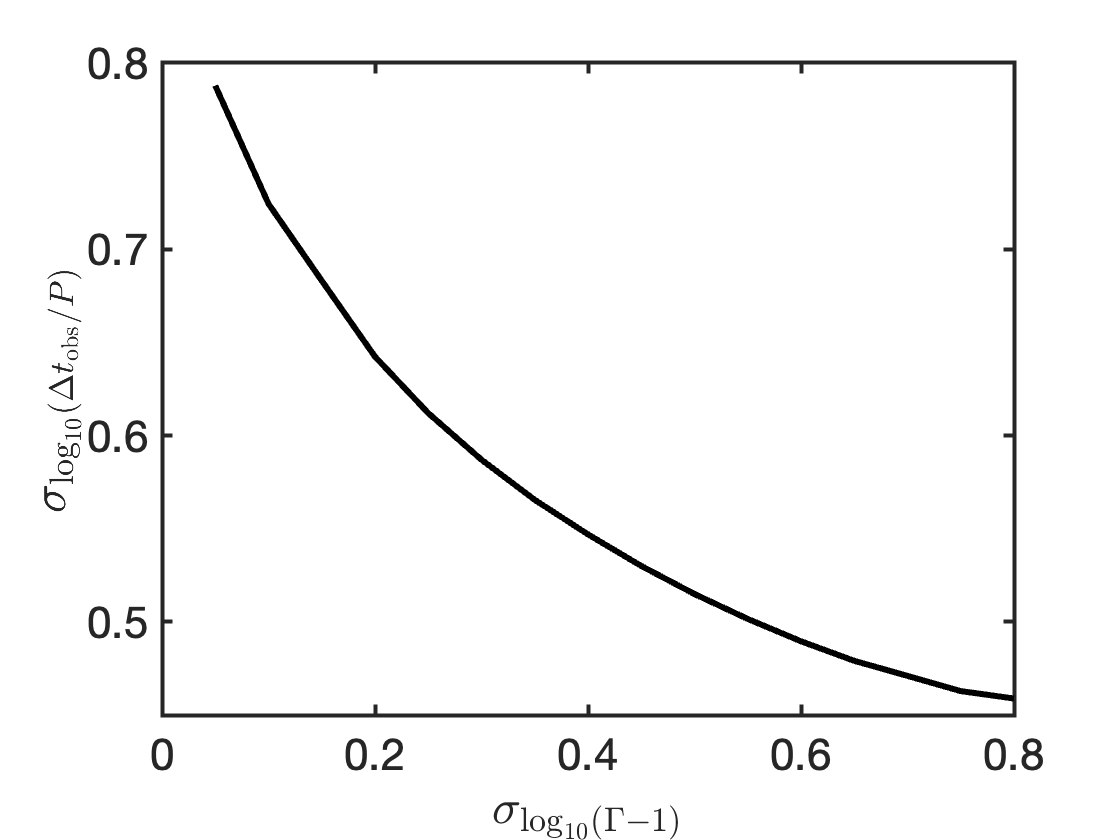}
		\caption{Top: Time interval up to next shock, o, (next shell emission, x,) as a function of observed shock (current shell emission) time. We assume here: $P=217\mbox{ ms}, \langle \log_{10}(\Gamma-1)\rangle=2,\sigma_{\log_{10}(\Gamma-1)}=0.3$, a total ejection time of $3$\,s as well as equal energy shells. Bottom: Standard deviation in arrival times as a function of standard deviation in shell LFs. We assume the median value of the LF distribution is $\Gamma=100$.}
		\label{fig:intshock}
	\end{figure}

	Consider a sequence of N shells ejected at constant intervals as dictated by the observed period: $t_{\rm ej,i}=(i-1)P$. To ensure internal collisions, these shells must be ejected at different LFs, $\Gamma_i$. In the central engine frame, the collision time between two shells is given by
	\begin{equation}
		t_{\rm coll,ji}=\frac{\beta_j t_{\rm ej,j}-\beta_i t_{\rm ej,i}}{\beta_j-\beta_i}
	\end{equation}
	Notice that the number of collisions depends sensitively on the distribution of LFs. Once two shells collide they form a combined shell. The properties of the combined shell depends both on the shells LF and their relative energies. Assuming (motivated by the comparable observed fluences in the individual peaks of FRB 20191221A) equal energy shells, the LF of the combined shell is 
	\begin{equation}
		\Gamma_{\rm com,ji}=\Gamma_i \left(\frac{2a_{\rm ji}^2}{a_{\rm ji}^2+1}\right)^{1/2}
	\end{equation}
	where $a_{\rm ji}\equiv \Gamma_{j}/\Gamma_i$.The combined shell can be considered to have an ``effective" ejection time 
	\begin{equation}
		t_{\rm ej,ji}=t_{\rm coll,ji}-\frac{\beta_i(t_{\rm coll,ji}-t_{\rm em,i})}{\beta_{\rm com,ji}}
	\end{equation}
	This is the time at which it would have been ejected if it were to reach the collision radius at the collision time with its current velocity.
	Photons emitted from this shell will reach the observer at a time
	\begin{equation}
		t_{\rm obs,ji}=t_{\rm coll,ji}-\beta_i(t_{\rm coll,ji}-t_{\rm em,i})
	\end{equation}
	After each such collisions the number of shells decreases by 1. The process continues either until all the shells have combined to a single shell or to a succession of shells with a distribution of LFs that is monotonically increasing with radius.
	In order to have collisions in the first place, it is necessary to have fluctuations in the LF distribution of the shells. At the same time, the process is non-linear and the number of collisions and observed time difference between successive collisions varies significantly, even with modest fluctuations in the LF distributions. An example of this is shown in figure \ref{fig:intshock} where we show the times of collisions for a specific realization of LFs in the shells. With $\sigma_{\log \Gamma}=0.3$ and equal energy shells, one finds an average value of $\sigma_{\log_{10}(\Delta t_{\rm obs})}\approx 0.6$.This is clearly in contradiction with the high accuracy within within which the periodicity is measured in FRB 20191221A, $\Delta P/P\!\sim\! 10^{-4}$. 
	
	\subsubsection{External shock between ejecta and magnetar wind}
	\label{sec:Extshock}
	We consider coherent radio emission in the shock interaction between a relativistic outflow and magnetar-wind; outflows interact with the wind from the magnetar that is released in between periodic launching of these outflows. The coherent emission we consider in this sub-section is the synchrotron maser process. However, constraints similar to what we find here would apply to other maser processes as well.
	
	Consider the flare-outflow and wind kinetic energy luminosities to be $L_{\rm f}$ and $L_{\rm w}$, which are given by
	\begin{equation}
		L_{\rm f} = 4\pi R^2 n'_F m c^3 \Gamma_{\rm F}^2, \quad\quad L_{\rm w} = 4\pi R^2 n'_w m c^3 \Gamma_{\rm w}^2,
	\end{equation}
	where $R$ is the distance from the NS, $n'_F$ \& $n'_W$ are particle densities for the outflow and wind in their rest frames, $\Gamma_{\rm F}$ \& $\Gamma_{\rm w}$ are LFs of the outflow and the wind respectively, and $m$ is e$^\pm$ mass. Pressure balance between the shocked outflow and the shocked wind across the surface of discontinuity gives $n'_F\gamma_1^2 \sim n'_w \gamma_2^2$, where $\gamma_1$ \& $\gamma_2$ are the LFs of the shocked outflow and wind wrt to the unshocked media ahead of the shock-fronts\footnote{This pressure balance condition assumes that both shocks are relativistic and the magnetization parameter ($\sigma$) for the flare outflow and the magnetar wind are of order unity or less, which has been shown by \cite{Metzger+19} to be favored for synchrotron maser in external shock for FRBs. The pressure balance condition can be easily generalized to consider $\sigma>1$. However, that does not change the main conclusion of this sub-section.}. The relative LF of the unshocked outflow wrt the unshocked wind is $2\gamma_1\gamma_2 =  \Gamma_{\rm F}/2\Gamma_{\rm w}$. Combining these equations yields 
	\begin{equation}
		\gamma_1 \approx \left[ {L_{\rm w}\over L_{\rm f}}\right]^{1/4} {\Gamma_{\rm F}\over 2\Gamma_{\rm w}},
	\end{equation}
	where we have implicitly assumed that $\Gamma_{\rm F}\gg\Gamma_{\rm w}$ (note that the outflow only decelerates significantly after it starts interacting with the wind). The deceleration radius for the outflow, i.e. the radius where the shock reaches the back end of the flare-outflow ejecta, is 
	\begin{equation}
		R_{\rm d} \approx c t_{\rm F} (\Gamma_{\rm F}^2/2\gamma_1^2) \approx {2c t_{\rm F}} \left[ {L_{\rm f}\over L_{\rm w}}\right]^{1/2} \Gamma_{\rm w}^2,
	\end{equation}
	where $t_{\rm F}$ is the duration of ejection of the flare-outflow in the neutron star rest frame, and $c\gamma_1^2/2\Gamma_{\rm F}^2$ is the speed of the shock front moving into the outflow; both of these are as viewed in the NS rest frame. We note that the deceleration time of the outflow in the observer frame is $t_{\rm d} \sim R_{\rm d}/(2 c \Gamma_{\rm F}^2/\gamma_1^2)\sim t_{\rm F}$. So, if outflows are ejected periodically, then their deceleration time and hence the radiation produced at $t_{\rm d}$ will be observed to be approximately periodic. We note that for self-consistency of the scenario discussed here, outflows should not collide at $r\lesssim R_{\rm d}$. This requires $L_{\rm f}/L_{\rm w} \gta (P/t_{\rm F})^2 \sim 10^4$; where $P=217$ ms is the time interval between outflow ejection, and $t_{\rm F} = 4$ ms is the width of the peaks of the radio lightcurve.  
	
	The maser frequency at the deceleration radius is approximately equal to $\omega_c \Gamma_{\rm F}/\gamma_1$, i.e. the product of the cyclotron frequency in the unshocked wind and the LF of the shocked outflow in the NS rest frame. The magnetic field in the unshocked-wind is $B_w' = (8\pi m c^2 n'_w \sigma_w)^{1/2}\propto \sigma_w^{1/2} t_{\rm F}^{-1} L_{\rm f}^{-1/2} L_{\rm w} \Gamma_{\rm w}^{-3}$; where $\sigma_w$ is the magnetization parameter for the wind. Therefore, the maser frequency in the observer frame at the deceleration time is 
	\begin{equation}
		\nu_{\rm obs}(t_{\rm d}) \propto \sigma_w^{1/2} t_{\rm F}^{-1} L_{\rm f}^{-1/4} L_{\rm w}^{3/4} \Gamma_{\rm w}^{-2}.
	\end{equation}
	
	We could require the frequency at deceleration to be in the CHIME band. This is possible provided that the right hand side of the above equation is constant. However, that imposes a stringent condition on the wind that $L_{\rm w} \sigma_w^{2/3} \Gamma_{\rm w}^{-8/3}$ should stay nearly exactly constant over the FRB duration of 2.5 s. Even a small change in the wind parameters will move the maser frequency at deceleration outside the CHIME band. The theoretical expectation is that wind luminosity should change substantially following a flare as the twisted (or distorted) magnetosphere has a different magnetic dipole moment.  If $\nu_{\rm obs}(t_{\rm d})$ is larger than 800 MHz, then the radiation will move into the CHIME band only at a later time. However, that will compromise the high degree of periodicity CHIME found for FRB 20191221A. We can quantify this from the dynamics of the blast wave. The relevant scalings are as follows: after deceleration, the LF of the shocked wind ($\Gamma$) declines with time in the observer frame ($t_{\rm ob}$) as $\Gamma \! \propto \! t_{\rm ob}^{-1/4}$, the radius of the shock front $R_{\rm sh}\! \sim \!2 c t_{\rm ob}\Gamma^2 \! \propto \! t_{\rm ob}^{1/2}$, and the synchrotron maser frequency $\nu_{\rm obs}\! \propto \! B_w' \Gamma \! \propto \! \Gamma\, R^{-1}\! \propto \!t_{\rm ob}^{-3/4}$. Let us consider a highly conservative case where the wind parameters do not change with time, however, flare-outflow luminosities reflect the variation of the observed light-curve (LC) peak amplitudes. If the external shock model could explain LC periodicity, surely it must for this case. Since, $\nu_{\rm obs}(t_{\rm d})\! \propto\! L_{\rm f}^{-1/4}$, the observer-frame time when the maser frequency enters the observing band deviates from the strict periodic launching of outflows by an amount of order $\delta t_{\rm ob}/t_{\rm d} \! \propto \! L_{\rm f}^{-1/3}$. The radio luminosity of FRB 20191221A varied by a factor $\sim \! 7$ during the 2.5 s that activity was recorded by CHIME. Thus, pulse arrival time should deviate from strict periodicity by $\sim \!2 t_{\rm d}\!\sim \!8$\,ms. However, according to CHIME, the positions of the peaks in the light-curve occurred at integral multiples of 216.8 ms to within 0.1 ms, which is at odds with the estimates presented here for even the most favorable case.
	
	\subsubsection{Refreshed external shock - multiple ejecta shells pilling up at external shock between ejecta and wind}
	
	The scenario considered in this section is that a flare-outflow interacts with magnetar wind, and the maser emission is produced in the transition layer of the freshly shocked wind close to the shock front. The decelerating shocked wind and the outflow system are hit from behind by the next flare-outflow launched by the NS thereby resulting in another peak in the LC, and so on. This scenario assumes that there is little wind in between successive flare episodes for the outflow to interact with. 
	
	The time interval between successive peaks in the LC is equal to P (ejection-period of outflows) plus at least the time it takes for shock wave from the collision with the newest flare-outflow to travel to the decelerating external shock front. The width of the peak of the LC would also be roughly equal to the shock crossing time. Since the successive collisions are taking place at larger and larger radii, the time interval between peaks would increase by an amount $\gta t_{\rm d}$ (which in this scenario is of order the spike duration $\sim 4$\, ms), and so too would the width of the peaks in the LC by a similar amount. These behaviors are inconsistent with CHIME data for FRB 20191221A. 
	
	\subsubsection{Forced reconnection}
	\label{sec:forcedrec}
	The forced reconnection model \citep{Lyubarsky2020,Mahlmann2022} considers a magnetic pulse that travels outwards from the magnetar, and compresses the plasma and magnetic field lines in the magnetar wind and forces oppositely aligned field lines (`stripes') of width $Pc$ in the unshocked wind to reconnect above the light-cylinder ($R_{\rm L}$); where $P$ is the magnetar's spin period.
	The energy carried by the magnetic pulse that forces reconnection is typically orders of magnitude larger than the energy that is needed to force reconnection in a single stripe. However, since the striped wind is moving at close to light-speed velocity, the next stripe will have propagated to distances $\sim \Gamma_{\rm w}^2 R_{\rm L}\gg R_{\rm L}$ before the pulse can catch up with it (where $\Gamma_{\rm w}$ is the LF of the wind). Depending on $\Gamma_{\rm w}$, such a large radius, may lie beyond the termination shock, inhibiting continued forced reconnection. Even if multiple stripes can be reconnected by a single pulse, their separation will not be periodic as the wind accelerates with radius (recall that the time difference between stripe reconnection events is $\Delta t\propto \Gamma_{\rm w}(r)^2$). Furthermore, the typical emitted frequency, which is roughly proportional to the cyclotron frequency, will have been reduced by many orders of magnitude between the first and subsequent events.
	
	This means that for forced reconnection to explain the observed periodicity in the arrival time of bursts requires multiple pulse ejections with periodic separations in their ejection time. This introduces various difficulties. First, similar to the case outlined in \S \ref{sec:Extshock}, if the luminosity of different pulses is variable (as observed for the spikes in FRB 20191221A) then this would change the maser frequency as $\nu\propto L_{\rm pul}^{5/8}$ which would cause a shift in the time that $\nu$ sweeps across the observed band and in turn destroy the high accuracy of the periodic signal.
	Second, the separation between consecutive pulse  ejections, $\Delta t_{\rm ej}$ should be $>P$ (the magnetar's period), in order for the magnetic stripe above the light-cylinder to be replenished before the next pulse comes through. This means the underlying magnetar's spin period should be shorter than the observed lightcurve periodicity in this scenario. However the former cannot be much smaller than the latter, since a magnetar with such a period would be very short lived, and its required age would become inconsistent with the observed DM of FRB 20191221A (see discussion of rotation powered models in \S \ref{sec:constraints}).
	Finally, even if magnetic flares are emitted at precisely periodic intervals, the location above the light-cylinder where the pulse first collides with the wind and starts the reconnection process will vary significantly from one event to the next (since the flare-period is uncorrelated with the magnetar's spin period), thus ruining the periodicity of the observed signal.
	
	\subsection{Energetic considerations}
	\label{sec:energyfar}
	In far-away models, the source of the observed FRB energy is magnetic. A common feature of far-away models, relying on synchrotron maser as the emission mechanism, is their low radiative efficiency, $\epsilon_{\rm rad}\!\approx \!10^{-6}\!-\!10^{-5}$ \citep{Metzger+19}\footnote{This efficiency includes both the efficiency of converting energy in the shock heated plasma to maser radiation as well as efficiency losses due to the fact that at the peak of the maser SED the optical depth is generally very large, and the observed peak is then shifted to higher frequencies where there is less power.}. We show here that there is an additional efficiency loss due to the fact that the escaping outflow can only propagate along open field lines. This means that there is a finite area on the surface of the NS around the magnetic pole from where the outflow can be launched, and thus only a small fraction of the magnetic field energy in the magnetosphere is tapped to power the outflow. We estimate the corresponding efficiency as $\epsilon_{\rm op}\!\sim \!\frac{1}{2}(\frac{R_{\rm c}}{R_0})^2$; where $R_{\rm c}$ is the polar cap radius and $R_0$ is the neutron star radius. The overall efficiency in a far-away model is then given by $\epsilon_{\rm tot}=\epsilon_{\rm op}\epsilon_{\rm rad}$.
	
	For the far-away model, the radius of the polar cap region from where magnetic field lines extend beyond the light cylinder is larger compared to an isolated NS due to the presence of an outflow.  In this case, $R_{\rm c}=R_0 \left(\frac{L}{B^2 R_0^2 c}\right)^{1/8}$\citep{Harding1999} where $L$ is the particle luminosity of the wind (which is assumed here to be greater than the spin-down luminosity, as is expected to be the case in far-away models). 
	This leads to
	\begin{equation}
		\epsilon_{\rm op}\approx 0.04 L_{\rm 47}^{1/4}B_{\rm 15}^{-1/2}.
	\end{equation}
	Note that this estimate is weakly dependent on the particle wind luminosity and decreases with increasing magnetic fields.
	All together, this suggests a canonical scaling of $\epsilon_{\rm tot}\sim 10^{-7}$ for the far-away model.

	Considering the observed fluence of FRB 20191221A, $\mathcal{F}$, and the DM based estimate of the distance, the required energy needed to power the burst in the far-away model can be constrained as
	\begin{equation}
		\label{eq:energyfar}
		E_{\rm req,far} \approx 4\pi \mathcal{F}\nu_p d_{\rm L}^2 \epsilon_{\rm tot}^{-1}\gtrsim 1.3\times10^{48}\,d_{\rm L,400}^2 \epsilon_{\rm tot,-7}^{-1} \mr{\,erg}
	\end{equation}
	where $\epsilon_{\rm tot,-7}\equiv\epsilon_{\rm tot}/10^{-7}$, $d_{\rm L,400}\equiv d_{\rm L}/400$Mpc and we have assumed that the emission is roughly isotropic, as is appropriate for a far-away model (see \S \ref{sec:intro}).
	
	It is useful to compare the required energy given by equation \ref{eq:energyfar} to the available magnetic energy in the NS, $E_{\rm mag}$ (. Requiring $E_{\rm mag}>E_{\rm req}$ we find
	\begin{equation}
		\label{eq:BE}
		B\gtrsim B_{\rm E,far}\equiv 2.5\times 10^{15} d_{\rm L,400} \epsilon_{\rm tot,-7}^{-1/2}  \mbox{ G}.
	\end{equation}
	where we approximate the volume of the NS as $4 \pi R_0^3/3$ to provide a rough approximation of the magnetic energy (in reality the field is not uniform in the star, but the volume weighted mean is likely dominated by a field that is of order the dipole field strength). Such a large surface dipole field strength is greater than observed for any of the Galactic magnetars, and would suggest a less common type of source for this scenario. While such a source is not ruled out by the observational requirements on the rate of FRB 20191221A-like sources, there are no known systems that harbor such fields.

	\begin{figure}
		\centering
		\includegraphics[width = 0.5\textwidth]{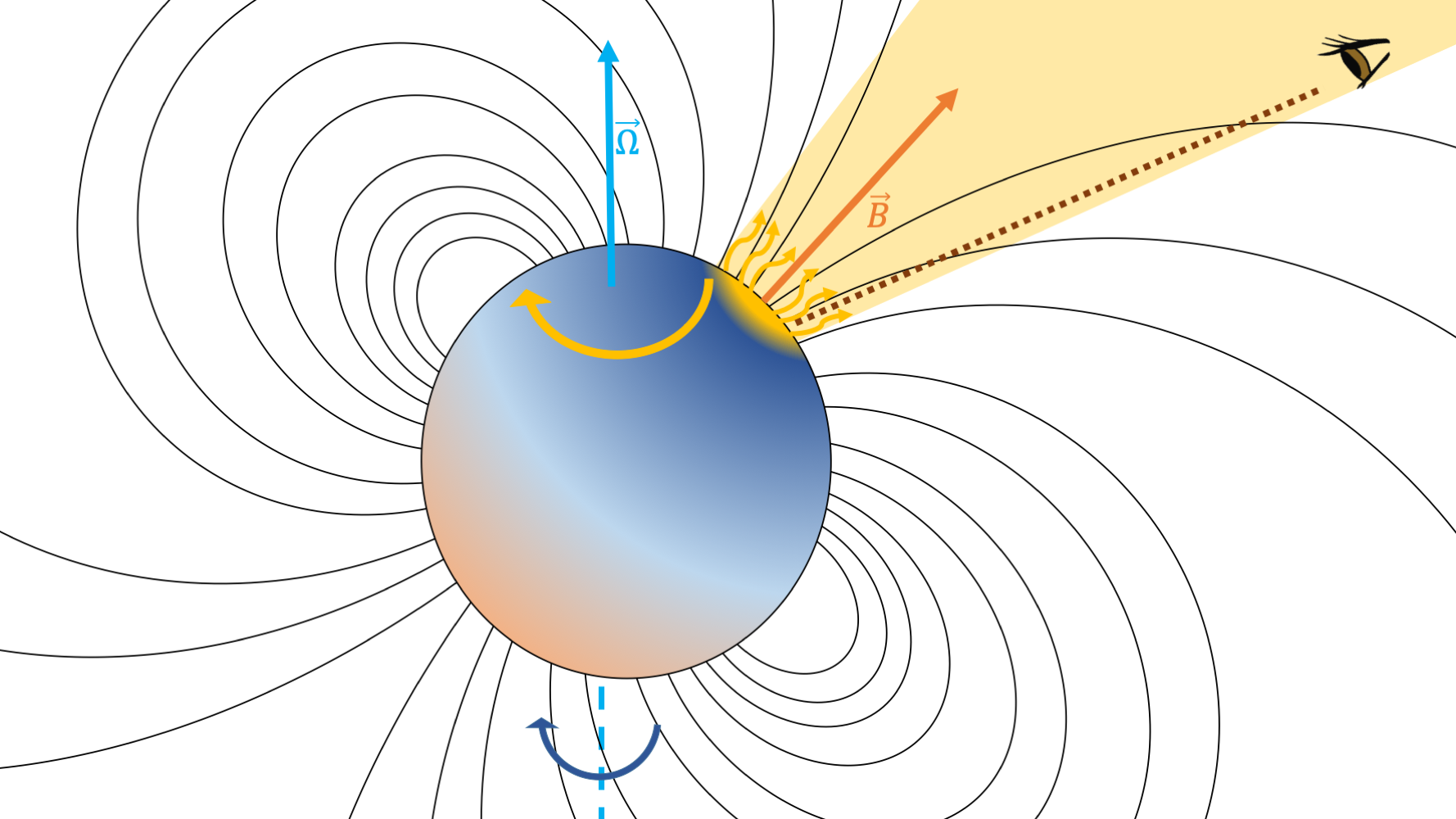}
		\caption{ Schematic figure of the rotating beam model.}
		\label{fig:rotatingbem}
	\end{figure}
	
	\section{Rotating beam magnetospheric model}
	\label{sec:constraints}
	Considering the difficulties with crustal oscillations as the origin of periodicity for FRB 20191221A (see \S \ref{sec:crustal}) and problems with the far-away models as emission sites for this event (see \S \ref{sec:FarAway}), we turn in this section to explore magnetospheric models involving a rotating beam in which the observed periodicity reflects the spin period (see figure \ref{fig:rotatingbem}).
	The energy source for such a model could be either the rotation of the neutron star or the magnetic energy reservoir. We explore the viability of both options below.
	A common feature of both variations of the rotating beam model is that the observed duty cycle can be naturally understood given the observed period. This is because the open field lines at radius $r\geq R_0$ are confined to an area of radius $r_c\approx r(r/R_{\rm LC})^{1/2}$; where $R_{\rm LC}=c/\Omega$ is the light-cylinder radius (at the NS surface, $r=R_0$ and hence $r_c=R_c$, i.e. the polar cap radius). The geometry of the sweeping radio beam -- assumed to be associated with open field lines -- then suggests that the duty cycle is
	\begin{equation}
		\label{eq:eta}
		\eta\sim \frac{2r_{\rm c}\sin \beta }{2\pi r\sin \alpha}\sim 0.01 r_{6}^{1/2} \frac{\sin \beta}{\sin \alpha}
	\end{equation}
	where $\alpha$ is the inclination of the magnetic axis relative to the spin and $\beta$ is the angle between the observer line-of sight and the magnetic dipole axis ($r_c \cos \beta$ is the impact parameter). With no special fine-tuning we would typically expect $\sin \beta/\sin \alpha \sim 1$. We see that equation \ref{eq:eta} naturally reproduces the observed duty cycle in FRB 20191221A's lightcurve as long as the radio emission is produced not too far away from the NS surface -- the distance of the emission region should be $r \lesssim 10 R_0$.
	The calculation above assumed a dipolar field structure. We note that as discussed in \S \ref{sec:energyfar}, in the presence of an outflow, the field line geometry is modified and, in particular, $R_{\rm op}$ decreases. As a result, $r_c\approx r (r/R_{\rm op})^{1/2}$ and $\eta$ increase for a given $r$. This means that the upper limit on $\eta$ based on the observed duty cycle becomes even more constraining for the allowed emission height in such a case as compared to the estimate quoted above for a dipole field (making the dipole assumption the conservative choice in this regard). Indeed, since $r_{\rm c} \propto L_{\rm p}^{1/8}$, this reasoning leads to an upper limit on the particle luminosity such that the observed value of $\eta$ may be reproduced for any value of $r\geq R_0$, $L_{\rm p}<6\times 10^{40} B_{15}^2 \mbox{erg s}^{-1}$.

	The emission is unlikely to be rotation powered as the estimates presented below show. It should be powered by magnetic energy, and the emission should persist for a few seconds. The duration of the emission episode is also highly constraining of possible scenarios, but we will not pursue that here.
	
	{\bf Rotation powered model}
	We consider first the rotation powered scenario. Considering the observed period of FRB 20191221A, the rotational energy of the NS is $E_{\rm rot}\sim \frac{1}{2} I \Omega^2\sim 5\times 10^{47}\mbox{ erg}$ which is easily large enough to account for the observed radio emission.
	However, more important for a rotationally powered source, is the available energy that can be extracted from rotation during the time of the FRB. In other words, we should compare the peak luminosity of the FRB to the spindown luminosity of the NS ($L_{\rm SD}$).
	\begin{equation}
		L_{\rm req} \approx 4\pi F_{\rm p} \nu_p d_{\rm L}^2 {f_{\rm b}\over \epsilon_{\rm tot}}\gtrsim 4\times10^{40}\,d_{\rm L,400}^2\left(\frac{f_{\rm b}}{0.02}\right) \left(\frac{10^{-1}}{\epsilon_{\rm tot}} \right)\mbox{ erg s}^{-1},    
	\end{equation}
	where we have normalized by a beaming fraction of $f_b$ of order the duty cycle, $\eta$, as required for a rotating beam model and by an efficiency of $\epsilon_{\rm tot}\sim 0.1$, which is comparable to the most powerful nano-shots observed in Galactic pulsars.
	We have also assumed that the bolometric flux is roughly $F\approx F_{\rm p}\nu_{\rm p}$ and $\nu_p=600$\,Mhz is the central frequency of the CHIME band \footnote{CHIME measures the flux only in the band $600-800$Mhz (at lower frequencies the flux drops below the noise). Nonetheless, the flux measured by CHIME for this burst is consistent with having a PL spectrum, which would suggest that $F\approx F_{\rm p}\nu_{\rm p}$ is a reasonable approximation. Furthermore, as mentioned in \S \ref{sec:intro}, the CHIME peak flux is strictly speaking just a lower limit}. The bolometric spin-down luminosity of a neutron star due to dipole radiation is (e.g., \citealt{ST1983})
	\begin{equation}
		L_{\rm SD} = 1.7\times 10^{41} B_{15}^2 P_{-1}^{-4} \,{\rm erg\, s^{-1}}\, ,
	\end{equation}
	Requiring $L_{\rm req}<L_{\rm SD}$ we constrain the magnetic field in the rotation powered scenario
	\begin{equation}
		B\gtrsim B_{\rm SD}\equiv 2.3\times 10^{15} \left(\frac{f_{\rm b}}{0.02}\right)^{1/2}\left(\frac{0.1}{\epsilon_{\rm tot}} \right)^{1/2} d_{\rm L,400} \mbox{ G}
		\label{B-sd}
	\end{equation}
	As mentioned before, no Galactic magnetar has dipole surface magnetic field this strong. Although that by itself cannot be used to discount the rotation powered model, there are other implications of the strong field that should be considered. The spin down time of a NS due to magnetic breaking is
	\begin{equation}
		\label{sec:tsd}
		\tau_{\rm SD} = 1.6\times 10^7 B_{15}^{-2} P_{-1}^2 \,{\rm s}\,
	\end{equation}
	or roughly 5 months for $B$ given by (\ref{B-sd}), and $p=217$ ms. If the NS were too young, then the DM associated with the shocked region between the SNR and the external medium would become prohibitively large \citep{PiroGaensler2018}. At ages younger compared to the Sedov Taylor time, $\tau_{\rm SD}\!\approx\! 10^3$\,yr, \cite{PiroGaensler2018} find that the shocked SNR dominates over the shocked external medium in its contribution to the DM. The contribution of the shocked SNR to the dispersion measure is estimated as $\mbox{DM}_{\rm SNR}\!\approx\! 50\, E_{\rm SN,51}^{-1/4}M_{\rm SN,1}^{3/4}n_0^{1/2}\tau_{\rm yr}^{-1/2}\mbox{pc cm}^{-3}$ \footnote{This estimate is likely to be conservative, as it considers contribution to the DM only from the ionized shock heated region of the ejecta. Radiation from the remnant (especially at younger ages) may lead to an ionization of some of the unshocked ejecta as well. This could potentially increase the SNR DM significantly as compared to the estimates by \cite{PiroGaensler2018}.}, where $M_{\rm SN,1}\!\equiv\! M_{\rm SN}/10M_{\odot}$ is the mass ejected in the SN, $E_{\rm SN,51}\!\equiv \! E_{\rm SN}/10^{51}$\,erg is the energy of the SN, $n_0\!\equiv \!n/1\mbox{ cm}^{-3}$ is the external medium density and $\tau_{\rm yr}\!\equiv \!\tau/\mbox{ yr}$ is the age. While the uncertainty in $E_{\rm SN},M_{\rm SN},n_0$, leads to a large uncertainty in $\mbox{DM}_{\rm SNR}$, we can nonetheless see that the measured value of $\mbox{DM}\!=\!368\mbox{pc cm}^{-3}$ suggests an age of $\tau>1$\,month at the very least.
	Using equation \ref{sec:tsd} we find an upper limit on the underlying NS magnetic field such that the age $\tau>1$\, month
	\begin{equation}
		\label{eq:Btau}
		B<B_{\tau}\approx 2.5\times 10^{15}\left(\frac{30\mbox{ days}}{\tau}\right)^{1/2}P_{-1}\mbox{ G }.
	\end{equation}
	This leaves a very narrow range of magnetic dipolar field strength at the NS surface, $2.3\times 10^{15}\mbox{ G}\!<\!B\!<\!5.4\times 10^{15}\mbox{ G}$, that is allowed by rotation powered models. It is inconceivable that such a young system is not a persistent radio source (due to shock interaction) detectable by CHIME or some other radio surveys.
	
	{\bf Magnetic powered model:}
	As rotation powered models are disfavoured, we turn next to magnetic powered models.
	An upper limit on the efficiency of magnetic powered models exists due to the fact FRBs can tap magnetic energy only in open field line region of the NS (see \S \ref{sec:energyfar}). Considering the size of the polar cap we find
	\begin{equation}
		\epsilon_{\rm op}\approx \frac{1}{2}\frac{\Omega R_0}{c}=1.2\times 10^{-3}P_{-1}.
	\end{equation}
	This represents an upper limit on the efficiency of magnetic powered, magnetospheric or near-field, models. Adopting this efficiency, and the angular size of rotating beam models (see above), our estimate for the required energy (equation \ref{eq:energyfar}) becomes
	\begin{equation}
		\label{eq:energynear}
		E_{\rm req,near} \approx 4\pi \mathcal{F}\nu_p d_{\rm L}^2 \epsilon_{\rm tot}^{-1}\gtrsim 2.6\times10^{42}\,d_{\rm L,400}^2 \left(\frac{f_{\rm b}}{0.02}\right) \epsilon_{\rm tot,-3}^{-1} \mr{\,erg}
	\end{equation}
	which leads to
	\begin{equation}
		B\gtrsim B_{\rm E,near}\equiv 3\times 10^{12}  \left(\frac{f_{\rm b}}{0.02}\right)^{1/2}\left(\frac{10^{-3}}{\epsilon_{\rm tot}} \right)^{1/2} d_{\rm L,400} \mbox{ G}.
	\end{equation}
	Comparing with the upper limit on the magnetic field due to spin-down age considerations ($B_{\tau}$; equation \ref{eq:Btau}) we see that there is ample parameter space for magnetically powered rotating beam models.
	
\begin{figure}
		\centering
		\includegraphics[width = 0.5\textwidth]{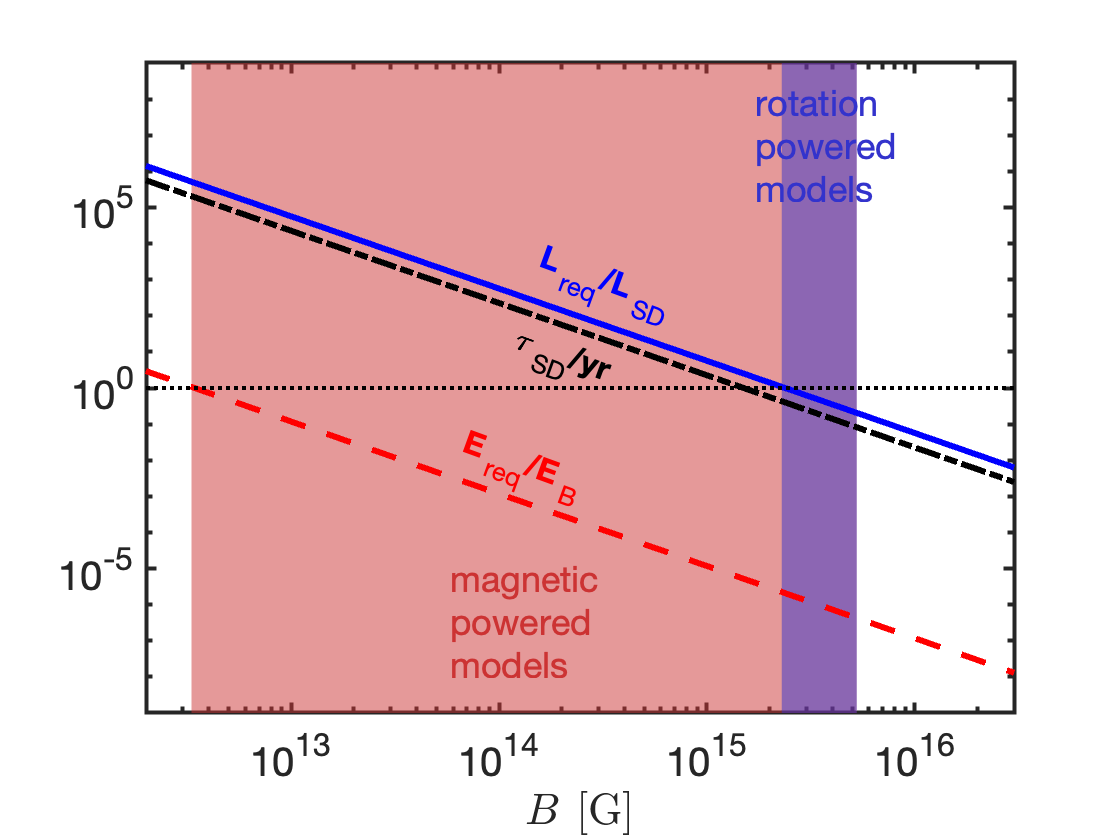}
		\caption{Constraints on the required magnetic field for magnetic and rotation powered models in the sweeping beam model. We take as a canonical value here $f_{\rm b}=0.02$ as well as $\epsilon_{\rm tot}=0.1$ for rotation powered models (and correspondingly for $L_{\rm req}/L_{\rm SD}$, depicted as a solid line) and $\epsilon_{\rm tot}=10^{-3}$ for the magnetic powered model (and correspondingly for $E_{\rm req}/E_{\rm B}$, depicted as a dashed line). Finally, a dot-dashed line depicts the spindown time in years. The allowed region for magnetic dominated models (for which $B$ is limited from below by the condition $E_{\rm req}/E_{\rm B}<1$ and from above by $\tau_{\rm SD}>1$\,month) is shown in red and the allowed region for rotation powered models (for which $B$ is limited from below by the condition $L_{\rm req}/L_{\rm SD}<1$ and from above by $\tau_{\rm SD}>1$\,month) is shown in blue.}
		\label{fig:Blims}
	\end{figure}

	Our results for rotation and magnetic powered rotating beam models are summarized in figure \ref{fig:Blims}. The figure demonstrates that while rotation powered models are confined to a narrow sliver of allowed $B$ values (which in any case corresponds to magnetar field strengths), the magnetic powered models can more naturally account for the observed properties of FRB 20191221A.

	\section{Discussion and conclusions}
	\label{sec:conc}
	We have considered various physical scenarios for the observations of FRB 20191221A, a non-repeating FRB with $216.8\pm 0.1$\,ms periodicity in the spacing between the peaks in its lightcurve. We have shown that NS models where the emission takes place beyond the light-cylinder (`far-away' models) do not explain the observed, highly regular, periodicity. Depending on the flavour of the far-away model, the high degree of stability in the observed periodicity requires either extreme fine tuning of the model parameters or predict a monotonic increase in the pulse duration over time, which is not observed. Energetic considerations further constrain the far away model and push the underlying NS to have very large magnetic field strengths, $B\gtrsim 2.5\times 10^{15}$\,G. In particular, the measured periodicity allows us to constrain the fraction of open field lines in the NS magnetosphere. This finite fraction corresponds to a finite area on the surface of the NS from which the outflow can be launched, and thus to a finite fraction of the magnetic field energy in the magnetosphere that can be tapped to power the outflow.
	This leads to a reduction of the overall efficiency of far-away models by a factor of $\sim 25$ beyond any constraints arising from radiative efficiency alone. 
	
	Several arguments have led us to prefer a spin period origin for the periodicity of FRB 20191221A rather than a period associated with crustal oscillations. These include the observed oscillation frequency in the radio lightcurve (5 Hz), which is low compared to the frequencies of numerous observed QPOs. The high degree of period stability over the 3s duration of the outburst is another reason for preferring NS spin period as the driver for the observed burst periodicity. Moreover, considering that NS crustal oscillations are seen at 100 Hz much more frequently than at 10 Hz whereas no FRB has shown any periodicity other than 20191221A at 5 Hz, is another reason against crustal oscillation mechanism. An additional clue is the presence of null periods. If crustal oscillations were to launch bursts and impose periodicity on the radio signal, then the presence of null periods pose an additional difficulty for the model.  Alternatively, if the origin of periodicity is the NS spin-period, the periodicity is decoupled from the origin of the outbursts. This makes null periods easy to understand as well as the fact that there are hundreds of hours of observation of the source with only one instance of activity \citep{subsecPCHIME} -- random bursting activity in the polar region lasting for about 3s. 
	
	These considerations suggest a rotating beam model for FRB 20191221A, in which the radiation beam periodically sweeps across the line of sight, as in pulsars. Considering the energetics of the system, we have shown that a NS powered by spin-down is only viable if the dipole surface magnetic field is $\gtrsim \! 2.5\!\times\! 10^{15}$\,G. At the same time, the measured DM of the system suggests that it cannot be a very young NS embedded in a SNR, leading to $B\!\lesssim \!5\!\times\! 10^{15}\,$G. This leaves a very narrow parameter space for pulsar-like rotation powered models. The large required magnetic field strength, suggests, instead, that the bursts may be more naturally powered by magnetic energy. Indeed, a magnetic powered rotating beam model is viable for a wide range of surface field strengths $3.5\!\times\! 10^{12}\!\lesssim \! B\!\lesssim\! 5\!\times \!10^{15}$\,G. We consider the latter as the most promising model for FRB 20191221A.
	
	Our model for the observed period for FRB 20191221A has implications for spin period distribution of FRB producing magnetars.
	In order to deduce an underlying period $P$ from an FRB lightcurve, its duration must be $t_{\rm FRB}\!\gg\! P$.
	Known Galactic magnetars have spin periods in the range $P_{\rm mag}\!\sim\! 2\!-\!12$\,s.
	There are as of yet no FRBs with burst duration $\!\gg\! P_{\rm mag}$ and thus no FRBs where such underlying spin periods could have been detected from the FRB lightcurve.
	However, a small subset of young magnetars, are expected to have much shorter spin periods. For a fixed dipole magnetic field (and while the spin evolution is dominated by dipole spin-down, as expected at $t\lesssim10^3\!-\!10^4$\,yrs), the fraction of magnetars with $P\!\ll \!P_{\rm mag}$ is $(P/P_{\rm mag})^2$ \footnote{This holds for $P\!\geq \!P_0$ where $P_0$ is the typical birth spin period of magnetars. The quoted scaling comes from the fact that so long as $B$ has not significantly decayed, $\tau_{\rm SD}\propto P^2$. Thus, for a fixed rate of magnetar formation, the fraction of the population that has a spin period $P$ is proportional to $P^2$.}.
	For example, for each magnetar with spin period $P\!=\! 200$\,ms there should be approximately $f_{10}\!\sim \!20^{-2}\!=\!1/400$ systems with $P\!=\!10$\,ms \footnote{This is possibly a lower limit, considering that FRBs may be more abundantly produced while the magnetar is younger and has a shorter spin-period (although the bulk of the magnetic energy dissipates on much longer timescales, multi-polar components may have shorter lifetimes associated with them).}.
	At the same time, the number of FRBs for which $10$\,ms periodicity could have been detected from the burst lightcurve is larger than FRBs for which $200$\,ms periodicity could have been detected (i.e. with durations of a few seconds, such as FRB 20191221A) by a factor of $N_{10}\!\sim\! 15$.
	All-together, we find that based on one detection of an FRB with $217$\,ms periodicity, the expected number of FRBs with 10\,ms periodicity in their lightcurves is  $f_{10}N_{10}\!\sim \!0.04\!\ll \!1$. In other words, the fact that no spin periods smaller than $200$\,ms have been seen in FRB lightcurves is consistent with magnetar spin evolutions. That being said, as the sample size of FRBs will increase, the rotating magnetar beam model implies that it should become possible to observe shorter underlying periods in FRB lightcurves. Once the FRB sample becomes much larger, this approach may even provide a constraint on the typical birth periods of magnetars, a property that is poorly determined from modeling of the observed Galactic magnetar population \citep{Beniamini2019}.

	\bigskip\bigskip
	\noindent {\bf ACKNOWLEDGEMENTS}
	
	\medskip
	PB was supported by a grant (no. 2020747) from the United States-Israel Binational Science Foundation (BSF), Jerusalem, Israel. PK's work was funded in part by an NSF grant AST-2009619.
	\medskip
	
	\noindent {\bf DATA AVAILABILITY}
	
	\medskip
	The code developed to perform calculations in this paper is available upon request.

\end{document}